\documentclass[twocolumn,showpacs,preprintnumbers,amsmath,amssymb]{revtex4}
\usepackage{graphics}
\usepackage{epsfig}
\usepackage{bm}

\begin{document}
\title{Tailoring magnetic and magnetocaloric properties of martensitic transitions in ferromagnetic Heusler alloys}
\author{Seda Aksoy, Thorsten Krenke\cite{AdressKrenke}, Mehmet Acet, Eberhard F. Wassermann}

\affiliation{Experimentalphysik, Universit\"at
Duisburg-Essen, D-47048 Duisburg, Germany}

\author{Xavier Moya, Llu\'{i}s Ma\~nosa, Antoni Planes}
\affiliation{Facultat de F\'isica, Departament d'Estructura i
Constituents de la Mat\`eria, Universitat de Barcelona, Diagonal
647, E-08028 Barcelona, Catalonia, Spain}
\date{\today}

\begin{abstract}
Ni$_{50}$Mn$_{34}$In$_{16}$ undergoes a martensitic transformation
around 250 K and exhibits a field induced reverse martensitic
transformation and substantial magnetocaloric effects. We
substitute small amounts Ga for In, which are isoelectronic, to
carry these technically important properties to close to room
temperature by shifting the martensitic transformation
temperature.
\end{abstract}


\maketitle

There is growing interest in searching for materials other than
Ni-Mn-Ga which may have interesting properties concerning
applications relevant to magnetic-field-induced strains. Such
search on Ni-Mn based Heusler systems has led to the observation
of giant magnetocaloric effects (MCE)
\cite{Hu00,Marcos02,Pareti03,Krenke05b,Han06,Sharma07}, large
strains related to field-induced transformations, and substantial
contribution to the understanding of martensitic transformations
in ferromagnetic Heusler materials. The valence electron
concentration ($e/a$) dependence of $M_s$ in NiMn$X$ is linear,
but with different slope for each $X$-species \cite{Krenke07c}.
Therefore, it should be possible to manipulate $M_s$ not only by
varying $e/a$, but also by holding $e/a$ constant and replacing
one $X$ species with another. In this manner one may have the
possibility of shifting and adjusting favorable features occurring
around the martensitic transformation of a particular alloy to
higher or lower temperatures. Ni$_{50}$Mn$_{34}$In$_{16}$
[$(e/a)\approx 7.87$] shows a field induced reverse martensitic
transformation at $M_s\approx 250$ K and associated with it, a
large field induced strain and a magnetocaloric effect
\cite{Krenke07,Moya07}. In view of technical interest, it would be
desirable to shift the transition temperature to around room
temperature without altering the favorable features. On the other
hand, in view of understanding the electronic properties of such
systems close to the martensitic transformation, it would be
interesting to understand to what extent the valence electron
concentration can be employed as a meaningful parameter. To test
this possibility, we substitute 2\% Ga for In in
Ni$_{50}$Mn$_{34}$In$_{16}$. From interpolation at constant
$(e/a)$, this amount of Ga is expected to shift $M_s$ to around
room temperature. We compare in this study the magnetic and
magnetocaloric properties of the isoelectronic compounds
Ni$_{50}$Mn$_{34}$In$_{16}$ \cite{Krenke07} and
Ni$_{50}$Mn$_{34}$In$_{14}$Ga$_{2}$ an discuss to what extent the
features around $M_s$ are preserved. The magnetocaloric properties
are studied from the entropy-change as well as from direct
temperature-change measurements.

\begin{figure}
\includegraphics[width=8cm]{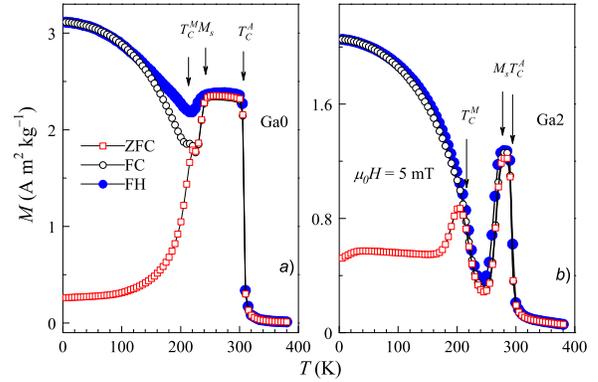}
\caption{\label{MT} (color online) (a) ZFC, FC, and FH $M(T)$ in 5
mT of a) Ga0 and b) Ga.}
\end{figure}

The samples were prepared by arc melting pure metals under argon
atmosphere. They were annealed at 1073 K for 2 hours and quenched
in ice-water. The compositions of the alloys were determined by
energy dispersive x-ray analysis.

Temperature dependent magnetization measurements $M(T)$ were
carried out in 5 mT in the temperature range $4<T<400$ K, and
magnetization isotherms $M(\mu_0 H)$ around the martensitic
transformation were obtained in magnetic-fields up to 5 T using a
superconducting quantum interference device magnetometer. The
entropy change $\Delta S$ was obtained from the magnetization
isotherms, and the direct temperature-change was measured with an
adiabatic magneto-calorimeter.

Figures \ref{MT}a and \ref{MT}b show $M(T)$ in 5 mT taken on a
zero-field-cooled (ZFC), field-cooled (FC), nd field-heated (FH)
sequence for Ga0 and Ga2 respectively. The curves corresponding to
the ZFC and FC states for both samples deviate below $T_C^M$,
whereas no appreciable deviation is found below $T_C^A$. The
deviation below $T_C^M$ is related to the anisotropy that develops
in the non-cubic martensitic phase of the alloys, so that cooling
in zero-field and cooling in finite field lead to different spin
configurations with different $M(T)$. For Ga0, $T_C^A\approx 308$
K, and this decreases to about 293 K for Ga2. On the other hand
$M_s$ increases from about 243 K for Ga0 to about 275 K for Ga2,
but the fundamental features of the curve remain similar.

\begin{figure}
\includegraphics[width=8cm]{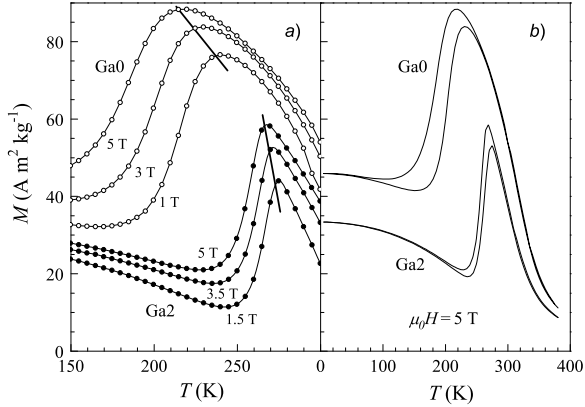}
\caption{\label{MTht} $M(T)$ for Ga0 and Ga2 in high fields. a)
Field-cooled $M(T)$ for Ga0 and Ga2. b) $M(T)$ for Ga0 and Ga2 in
the FC and FH states. The thermal hysteresis is broader in Ga0.}
\end{figure}

$M(T)$ has been also measured in several fields $\mu_0 H\geq 1$ T
to compare the field rate of shift of $M_s$, $dM_s/dH$, of both
samples. The results are shown in Fig. \ref{MTht}a. The heavy
lines are drawn through the points joining the onset of decrease
in $M(T)$ with decreasing temperature. These mark $M_s$ for each
measuring field. The slope of these lines at these magnetic-field
ranges give $dM_s/dH\approx -6$ KT$^{-1}$ and $dM_s/dH\approx -2$
KT$^{-1}$ for Ga0 and Ga2 respectively. $M(T)$ curves in 5 T for
the FC and FH states are shown in Fig. \ref{MTht}b. The thermal
hysteresis for Ga2 narrows with respect to that of Ga0.
Furthermore, it is also seen that the magnetization in the
martensitic state decreases when Ga is added.

\begin{figure}
\includegraphics[width=8cm]{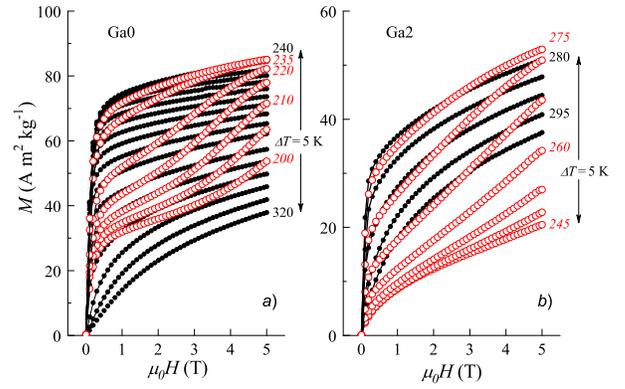}
\caption{\label{MH} (color online) magnetic-field dependence of
the magnetization for a) Ga0 and b) Ga2. Open circles (red) and
filled circles are data for $T<M_s$ and $T>M_s$ respectively.}
\end{figure}

The magnetization isotherms in the vicinity of $M_s$ in Figs.
\ref{MH}a and \ref{MH}b show that the overall agnetization is
lower in Ga2 than in Ga0. The data shown with open circles in both
figures correspond to $M(\mu_0 H)$ for $T<M_s$ (values printed in
italic), and the filled circles correspond to $T>M_s$. The
metamagnetic-like character of the feature in $M(\mu_0 H)$ at
temperatures $T<M_s$ is associated with a field-induced reverse
martensitic transformation. $M(\mu_0 H)$ initially increases with
increasing field with decreasing curvature, until it reaches an
inflection point at a field corresponding to the onset of the
field-induced transformation. Above this point, $M(\mu_0 H)$
begins to increase faster with increasing magnetic-field. For Ga2,
the field-induced transformation begins to take place at lower
fields than those needed for Ga0, so that the steep rise in
$M(\mu_0 H)$ begins already below 1 T. The narrower hysteresis in
$M(T)$ for Ga2 compared to broader hysteresis for Ga0 is the cause
for the lower threshold of the transformation in Ga0.

\begin{figure}
\includegraphics[width=8cm]{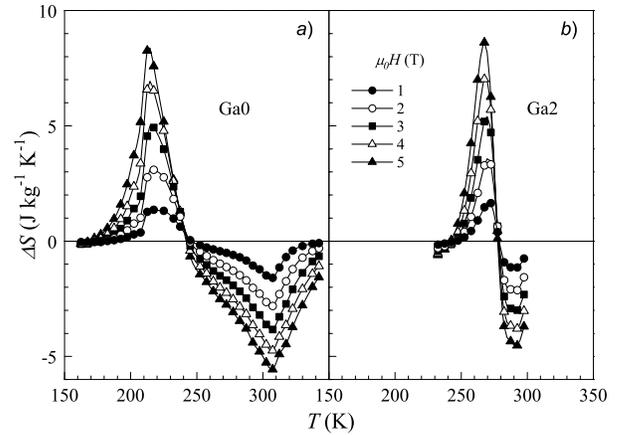}
\caption{\label{delS} Temperature dependence of the entropy change
around $M_s$ and $T_C^A$ for a) Ga0 and b) Ga2.}
\end{figure}

\begin{figure}
\includegraphics[width=8cm]{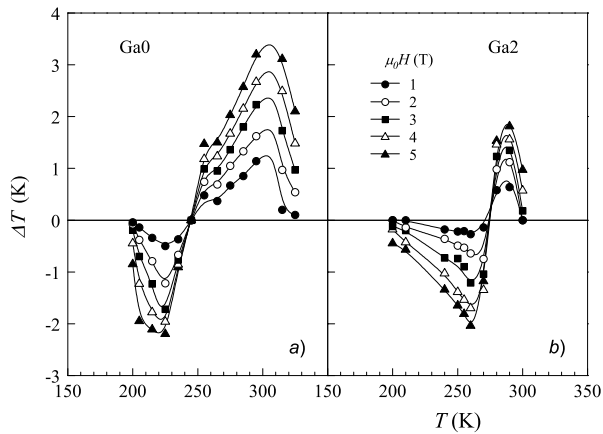}
\caption{\label{delT} Temperature dependence of the measured
temperature change around $M_s$ and $T_C^A$ a) Ga0 and b) Ga2.}
\end{figure}

Using the data in Fig. \ref{MH}, the field induced entropy change
$\Delta S$ is determined by ntegrating numerically $\Delta S(T,H)
= \mu_0 \int^H_{0} ({\partial M}/{\partial T})_H dH$. $\Delta
S(T,H)$ for Ga0 and Ga2 is shown in Figs. \ref{delS}a and
\ref{delS}b. For both samples $\Delta S(T,H)$ is positive below
$M_s$ (inverse MCE) and negative around $T_C^A$ (conventional MCE)
with the crossover taking place at the temperature corresponding
to $M_s$ determined from Figs. \ref{MT}a and \ref{MT}b. The
magnitude of the entropy change below $M_s$ remain nearly
unchanged for both samples, with a maximum value of 8
Jkg$^{-1}$K$^{-1}$. Above $M_s$, $\Delta S$ of Ga0 reaches a
slightly higher value than that of Ga2, both being about $-$5
Jkg$^{-1}$K$^{-1}$ under 5 T. As expected, both samples cool on
applying a magnetic-field below $M_s$ and heat on applying a field
around $T_C$ as seen from the results of the direct magnetocaloric
measurements in Figs. \ref{delT}a and \ref{delT}b for both
samples. The maxima in $\Delta T$ below $M_s$ are nearly the same
for both samples reaching a value of $-$2 K in 5 T. Around
$T_C^A$, the maximum value is about 3.5 K for Ga0 and is slightly
larger than 2 K for Ga2. This difference is consistent with the
difference in $\Delta S$ above $M_s$.

Investigations on quaternary Heusler-based systems have been
undertaken previously both to improve material properties and to
examine the interplay between magnetic and structural properties
around the martensitic transformation
\cite{Khan04,Khan05,Gao06,Xuan07,Krenke07b,Kainuma06}. We provide
in this study a method based on the varying $e/a$ dependence of
$M_s$ for different group III-V elements, by which $M_s$ can be
shifted so that favorable properties of a particular alloy can be
brought to a desired temperature. Presently, this desired
temperature is limited to below room temperature, since $T_C^A$ is
limited to about 300-350 K in NiMn-based Heusler alloys.
Nevertheless, we find indeed that at constant $e/a$, it is
possible to preserve at a high temperature a favorable MCE
property of Ni$_{50}$Mn$_{34}$In$_{16}$ occurring at a low
temperature by substituting an element isoelectronic to In, namely
Ga. The next step would be to devise a method to manipulate $T_C$,
such as that involving the addition of small amounts of Co. Some
work already gives evidence that replacing Ni with small amounts
of Co tends to increase $T_C^A$ \cite{Yu07}. This would be
particularly interesting, e.g., in a sample with about 4 at\% Ga,
where $M_s$ lies already within the paramagnetic regime of both
austenite and martensite phases. By adding Co, such a system would
regain its ferromagnetism in the austenitic state.

We find in the present studies that the maximum absolute values of
$\Delta S$ and $\Delta T$ on both sides of $M_s$ for Ga0 and Ga2
are nearly the same. $dM_s/dH$ is smaller for Ga2 than for Ga0
meaning that the MCE on Ga2 should be smaller. It appears that the
narrower temperature hysteresis for Ga2 with respect to that of
Ga0 facilitates the field induced reverse transformation. As
discussed in previous studies, the narrow hysteresis is favorable
for a large MCE, and much effort is invested in reducing
hysteresis-losses \cite{Gschneider05,Provenzano04}. The reduced
hysteresis in Ga2 compensates for its lower $dM_s/dH$ as compared
to that of Ga0. As can be seen from the fast rise of the
magnetization with increasing magnetic-field already in low-fields
in Fig. \ref{MH}, a lower threshold field for Ga2 is required than
that for Ga0 to induce a transformation with an external field.

This work was supported by Deutsche Forschungsgemeinschaft (GK 277
and SPP 1239) and CICyT (Spain), project MAT2007-61200. XM
acknowledges support from DGICyT (Spain).


\begin{thebibliography}{99}

\bibitem[*]{AdressKrenke} Present address: ThyssenKrupp Electrical Steel, Kurt-Schumacher-Str. 95, D-45881 Gelsenkirchen, Germany

\bibitem{Hu00} F. Hu, B. Shen, J. Sun, Appl. Phys. Lett.
\textbf{76}, 3460 (2000).

\bibitem{Marcos02} J. Marcos, A. Planes, L. Ma\~nosa, F. Casanova, X. Batlle, A. Labarta, and B. Mart\'{\i}nez,
 Phys. Rev. B \textbf{66}, 224413 (2002).

\bibitem{Pareti03} L. Pareti, M. Solzi, F. Albertini, A. Paoluzi, Eur. Phys. J. B, \textbf{32}, 303 (2003).

\bibitem{Krenke05b} T. Krenke, M. Acet, E. F. Wassermann, X. Moya, L. Ma\~nosa, A. Planes, Nature Materials \textbf{4}, 450 (2005).

\bibitem{Han06} Z. D. Han, D. H. Wang, C. L. Zhang, S. L. Tang, B. X. Gu, Y. W. Du, Appl. Phys. Lett. \textbf{89}, 182507 (2006).

\bibitem{Sharma07} V. K. Sharma, M. K. Chatttopadhyay, S. B. Roy, J. Phys. D: Appl. Phys \textbf{40}, 1869 (2007).

\bibitem{Krenke07c}T. Krenke, X. Moya, S. Aksoy, M. Acet, P. Entel, Ll. Ma\~nosa, A. Planes, Y. Elerman, A. Y\"ucel, E.F. Wassermann, J. Magn. Magn. Mater. \textbf{310} 2788 (2007).

\bibitem{Krenke07} T. Krenke, E. Duman, M. Acet, E. F. Wassermann, X. Moya, L. Ma\~nosa, A. Planes, E. Suard, B. Ouladdiaf, Phys. Rev. B. \textbf{75}, 104414 (2007).

\bibitem{Moya07}X. Moya, L. Ma\~nosa, A. Planes, S. Aksoy, T. Krenke, M. Acet, E. F. Wassermann, Phys. Rev. B \textbf{75}, 184412 (2007).

\bibitem{Khan04} M. Khan, I. Dubenko, S. Stadler, N. Ali, J. Phys.: Condens. Matter. \textbf{16}, 5259 (2004).

\bibitem{Khan05} M. Khan, I. Dubenko, S. Stadler, N. Ali, J. Appl. Phys. \textbf{97}, 10M304 (2005).

\bibitem{Gao06} L. Gao, W. Cai, A. L. Liu, L. C. Zhao, J. Alloy. Comp. \textbf{425}, 314 (2006).

\bibitem{Kainuma06} R. Kainuma, Y. Imano, W. Ito, , Y. Imano, W. Ito, Y. Sutou, H. Morito, S. Okamoto, O. Kitakami, K. Oikawa, A. Fujita, T. Kanomata, and K. Ishida, Nature \textbf{439}, 957 (2006).

\bibitem{Xuan07} H. C. Xuan, D. H. Wang, C. L. Zhang, Z. D. Han, H. S. Liu, B. X. Gu, Y. W. Du, Sol. State Comm. \textbf{142}, 591 (2007).

\bibitem{Krenke07b} T. Krenke, E. Duman, M. Acet, E. F. Wassermann, X. Moya, L. Ma\~nosa, A. Planes, J. Appl. Phys. \textbf{102}, 033903 (2007).

\bibitem{Yu07} S. Y. Yu, L. Ma, G. D. Liu, Z. H. Liu, J. L. Chen, Z. X. Cao, G. H. Wu, B. Zhang, X. X. Zhang, Appl. Phys. Lett. \textbf{90}, 242501 (2007).

\bibitem{Provenzano04} V. Provenzano, A. J. Shapiro, and R. D. Shull, Nature \textbf{429}, 853 (2004).

\bibitem{Gschneider05} K. A. Gschneidner Jr., V. K. Pecharsky, and A. O. Tsokol, Rep. Prog. Phys. \textbf{68}, 1479 (2005).

\end{thebibliography}
\end{document}